\documentclass[aps,prd,floatfix,showpacs,11pt]{revtex4}

\usepackage{amsmath,amsfonts}
\usepackage[dvips]{graphicx}
\usepackage{epsfig}

\newcommand{\ud}{\mathrm{d}}
\newcommand{\ve}{\varepsilon}

\begin{document}

\title{Non-minimally coupled scalar field cosmology on the phase plane}

\author{Orest Hrycyna} 
\email{hrycyna@kul.lublin.pl}
\affiliation{Department of Theoretical Physics, Faculty of Philosophy, 
The John Paul II Catholic University of Lublin, Al. Rac{\l}awickie 14, 20-950
Lublin, Poland}
\author{Marek Szyd{\l}owski}
\email{uoszydlo@cyf-kr.edu.pl}
\affiliation{Dipartimento di Fisica Nucleare e Teorica, Universit{\`a} degli studi
di Pavia, via A. Bassi 6, I-27100 Pavia, Italy}
\affiliation{Mark Kac Complex Systems Research Centre, Jagiellonian University,
Reymonta 4, 30-059 Krak{\'o}w, Poland}

\date{\today}

\begin{abstract}
In this publication we investigate dynamics of a flat FRW cosmological model
with a non-minimally coupled scalar field with the coupling term 
$\xi R \psi^{2}$ in the scalar field action. The quadratic potential function
$V(\psi)\propto \psi^{2}$ is assumed. All the evolutional paths are visualized
and classified in the phase plane, at which the parameter of non-minimal
coupling $\xi$ plays the role of a control parameter. The fragility of global
dynamics with respect to changes of the coupling constant is studied in details.
We find that the future big rip singularity appearing in the phantom scalar
field cosmological models can be avoided due to non-minimal coupling constant
effects. We have shown the
existence of a finite scale factor singular point (future or past) where 
the Hubble function as well as its first cosmological time derivative diverge.
\end{abstract}


\maketitle

\section{Introduction}
\label{sec:1}
Recently scalar fields have played a very important role in cosmology. They are
used in many phenomenological models like quintessence \cite{Wetterich:1987fm,
Ratra:1987rm}. Scalar fields are also very important in description of dynamics
in the loop quantum cosmology, which base on the background independent theory
without the canonical notion of time. In this theory one scalar field is chosen
as an internal clock for other fields \cite{Ashtekar:2006rx}. The scalar fields
with a potential function are also very important in modelling of inflation. For
example a scalar field with the simplest quadratic potential function was
assumed in Linde's conception of chaotic inflation \cite{Linde:1983gd}. The 5
years of WMAP observations \cite{Komatsu:2008hk} rejected many inflationary
scenarios (potential functions $V(\psi)$), while models with a simple quadratic
potential are admitted on the $1\sigma$ confidence level.

In the standard quintessence energy density of the minimally coupled to gravity
scalar field mimics the effective cosmological constant. Of course the detailed
evolution is dependent upon a specific form of the potential $V(\psi)$ but the 
$\psi^{2}$ contribution can be treated as a leading order term of expansion of
the potential function.

We extend the quintessence scenario by incorporating the non-minimal coupling
constant \cite{Faraoni:2006ik, Faraoni:2000gx, Hrycyna:2007mq, Hrycyna:2007gd,
Szydlowski:2008zza}. In this paper we present a phase space analysis of the
evolution of a spatially flat Friedmann-Robertson-Walker (FRW) universe 
containing a non-minimally coupled to gravity scalar field, both canonical and 
phantom, with the simplest form of quadratic potential function. The similar 
analysis for the case of minimally coupled scalar field was performed in 
\cite{UrenaLopez:2007vz}. We extend this analysis on models with a non-zero 
coupling constant $\xi$ which plays the role of a control parameter for an 
autonomous dynamical system on the phase plane. Therefore the location of 
fixed points (physically representing asymptotic states of the system) as well 
as their character depends upon the value of $\xi$. The values of parameter
$\xi$ for which the global dynamics changes dramatically are called bifurcation
values. In our previous paper \cite{Szydlowski:2006qn}, in case of conformal
coupling and quadratic potential function we have
shown that phantom cosmology can be treated as a simple model with a scattering
of trajectories whose character depends crucially on the sign of the potential
function. We also demonstrated that there is a possibility
of chaotic behavior in the flat Universe with a conformally coupled phantom field in the system considered on the non-zero energy level.

The minimally coupled scalar field endowed with a quadratic potential
function has a strong motivation in inflationary models and its generalizations
with a simple non-minimal coupling term $\xi R \psi^{2}$ have been studied
\cite{Park:2008hz} in the context of the origin of the canonical inflaton field
itself. The physical motivation to investigate a non-minimally coupled scalar
field cosmological models could be possible application of this models to
inflationary cosmology or to the present dark energy (see for example 
\cite{Spokoiny:1984bd, Salopek:1988qh, Fakir:1992cg, Barvinsky:1994hx,
Barvinsky:1998rn, Uzan:1999ch, Chiba:1999wt, Amendola:1999qq, Perrotta:1999am,
Holden:1999hm, Bartolo:1999sq, Boisseau:2000pr, Gannouji:2006jm, Carloni:2007eu,
Bezrukov:2007ep, Barvinsky:2008ia}).

The coupling constant $\xi$ is a free parameter of the theory which should be
estimated from the observational data. Recently we have shown that distant
supernovae can be used to estimate the value of this parameter
\cite{Szydlowski:2008zza}.

The main advantage of dynamical system analysis is that we can visualize all the
trajectories of the system admissible for all initial conditions. Therefore one
can classify generic routes to the accelerating phase (the de Sitter attractor 
where $p_{\psi}=-\rho_{\psi}$). This attractor corresponds to the model with the
cosmological constant.

The paper has a following organization: in section \ref{sec:2} we reduce
dynamics to the form of an autonomous dynamical system which describes both
canonical and phantom scalar field models. Section \ref{sec:3} is devoted 
to a detailed analysis of the phase portraits for different values of the 
parameter $\xi$. In this section we also discuss the change of evolutionary 
scenarios upon the value of parameter $\xi$. 

\section{Non-minimally coupled scalar field cosmologies as a dynamical system}
\label{sec:2}

We assume the flat model with the FRW geometry, i.e. the line element has the
form
\begin{equation}
\ud s^{2} = -\ud t^{2} + a^{2}(t)[\ud r^{2} + r^{2}(\ud\theta^{2}
+ \sin^{2}{\theta}\ud\varphi^{2})],
\label{eq:1} 
\end{equation}
where $0 \leq \varphi \leq 2\pi$, $0 \leq \theta \leq \pi$ and $0 \leq r \leq
\infty$ are the comoving coordinates and $t$ stands for the cosmological time.

It is also assumed that a source of gravity is a scalar field $\psi$ with a
generic coupling to gravity. The gravitational dynamics is described by the
standard Einstein-Hilbert action
\begin{equation}
S_{g}=\frac{1}{2}m_{p}^{2}\int \ud^{4}x \sqrt{-g}R,
\label{eq:3}
\end{equation}
the action for the matter source is
\begin{equation}
S_{\psi} = -\frac{1}{2}\int \ud^{4}x
\sqrt{-g}\left[\ve\big(g^{\mu\nu}\psi_{\mu}\psi_{\nu} + \xi R\psi^{2}\big) + 2
V(\psi)\right].
\label{eq:4}
\end{equation}
where $m_{Pl}^{2}=1/(8\pi G)=1/\kappa$ and $R=6(\ddot{a}/a+\dot{a}^{2}/a^{2})$
and $\ve=+1,-1$ corresponds to the scalar field and the phantom scalar field,
respectively. For simplicity and without lost of generality
we will assume $4\pi G/3=1$ which corresponds to $\kappa=6$.

After dropping the full derivatives with respect to time we obtain the
dynamical equation for scalar field from variation
$\delta(S_{g}+S_{\psi})/\delta \psi=0$
\begin{equation}
\ddot{\psi}+3H\dot{\psi} +\xi R \psi + \ve V'(\psi)= 0.
\label{eq:7}
\end{equation}
as well as the energy conservation condition from variation
$\delta(S_{g}+S_{\psi})/\delta g=0$ 
\begin{equation}
\mathcal{E} = \ve\frac{1}{2}\dot{\psi}^{2} + \ve 3\xi H^{2}\psi^{2} + 
\ve 3\xi H (\psi^{2})\dot{} + V(\psi) - \frac{3}{\kappa}H^{2}
\end{equation}
If we include other forms of matter this condition can be expressed as
\begin{equation}
\frac{3}{\kappa}H^{2} = \rho_{\psi} + \rho_{r} + \rho_{m}
\end{equation}
where $\rho_{r}$ and $\rho_{m}$ are energy densities of radiation and matter,
respectively. It can be shown that for any value of $\xi$ scalar field behaves
like a perfect fluid with energy density $\rho_{\psi}$ and pressure $p_{\psi}$

\begin{subequations}
\begin{eqnarray}
\rho_{\psi}&=& \varepsilon\frac{1}{2}\dot{\psi}^{2} + V(\psi) +
\varepsilon\xi\big[ 3H(\psi^{2})\dot{} + 3H^{2}\psi^{2}\big],\\
p_{\psi}&=& \varepsilon\frac{1}{2}\dot{\psi}^{2} - V(\psi) -
\varepsilon\xi\big[2H(\psi^{2})\dot{} + (\psi^{2})\ddot{} +
(2\dot{H}+3H^{2})\psi^{2}\big].
\end{eqnarray}
\end{subequations}

Changing the dynamical variables according to the relation
$$
\dot{\psi} = \frac{\ud\psi}{\ud t} = \frac{\dot{a}}{a} \frac{\ud\psi}{\ud\ln{a}}
= H \psi'
$$
we can express the Hubble function as
\begin{equation}
\label{hub}
H^{2} = 2\frac{V(\psi)+\rho_{r}+\rho_{m}}{\frac{6}{\kappa} -
\varepsilon\big[(1-6\xi)\psi'^{2} + 6\xi(\psi'+\psi)^{2}\big]}.
\end{equation}
The denominator of (\ref{hub}) equal zero denotes a line in the phase space
of singularities of the Hubble function which separates the phase space in two
regions one physical $H^{2}>0$, and the second one nonphysical $H^{2}<0$. It
does not depend on the form of the potential function but only on a value of
the coupling constant. 

The Euler-Lagrange equations for the system under consideration are given in
the form
\begin{subequations}
\label{eq:9}
\begin{align}
 a^{2}\frac{\ud^{2}\psi}{\ud\eta^{2}} + 6\xi a\psi\frac{\ud^{2}a}{\ud\eta^{2}} 
& =  -2a\frac{\ud a}{\ud\eta}\frac{\ud\psi}{\ud\eta} - \ve a^{4} V'(\psi), \\
 \frac{6}{\kappa}\frac{\ud^{2}a}{\ud\eta^{2}}(1-\ve\kappa\xi\psi^{2}) -
\ve6\xi a\psi\frac{\ud^{2}\psi}{\ud\eta^{2}}  & = -\ve a
(1-6\xi)\Big(\frac{\ud\psi}{\ud\eta}\Big)^{2} + \ve12\xi\psi\frac{\ud
a}{\ud\eta}\frac{\ud\psi}{\ud\eta} + 4a^{3}V(\psi)+\rho_{m,0}.
\end{align}
\end{subequations}
where $\eta$ stands for the conformal time, $\ud\eta=\ud t/a$. 

After the elimination of the scale factor and its derivative from
system~(\ref{eq:9}) we obtain the condition
\begin{equation}
\begin{split}
\big(\psi''+\psi'\big)\Big(\frac{6}{\kappa}-\ve6\xi(1-6\xi)\psi^{2}\Big) -
\ve\psi'^{2}(1-6\xi)\big(\psi'+6\xi\psi\big) + \\ + \frac{1}{H^{2}}\bigg\{
\ve\frac{6}{\kappa}V'(\psi)\big(1-\ve\kappa\xi\psi(\psi'+\psi)\big) +
\big(4V(\psi)+\rho_{m}\big)\big(\psi'+6\xi\psi\big)\bigg\} = 0.
\label{eq:10}
\end{split}
\end{equation}
where the prime denotes differentiation with respect to the natural logarithm of
the scale factor.

Now we can simply present equation (\ref{eq:10}) in the form of the autonomous
dynamical system
\begin{subequations}
\begin{align}
&\psi'= y, \\
&y'= - y + 
\frac{\ve y^{2}(1-6\xi)\big(y+6\xi\psi\big)}{\Big(\frac{6}{\kappa} -
\ve6\xi(1-6\xi)\psi^{2}\Big)} - \frac{1}{2}\frac{\Big(\frac{6}{\kappa} -
\ve\big[(1-6\xi)y^{2} + 6\xi(y+\psi)^{2}\big]\Big)}{\Big(\frac{6}{\kappa} -
\ve6\xi(1-6\xi)\psi^{2}\Big)\Big(V(\psi)+\rho_{r}+\rho_{m}\Big)} 
\times \nonumber \\ &
\times\Big(\ve\frac{6}{\kappa}V'(\psi)\big(1-\ve\kappa\xi\psi(y+\psi)\big) +
\big(4V(\psi)+\rho_{m}\big)\big(y+6\xi\psi\big)\Big).
\end{align}
\end{subequations}
In what follows we will assume a quadratic potential function 
$V(\psi)=1/2 m^{2}\psi^{2}$, and that there is no other form of matter than the
scalar field, i.e that $\rho_{r}=\rho_{m}=0$. It is easy to notice that in this
case the dynamics does not depend on the change of a sign of the potential
$V(\psi)\to-V(\psi)$.
Finally, the dynamical system is in the form
\begin{subequations}
\label{syst}
\begin{align}
&\alpha\frac{\ud \psi}{\ud \sigma}=y\psi\Big(\frac{6}{\kappa} - \ve6\xi(1-6\xi)\psi^{2}\Big), \\
&\alpha\frac{\ud y}{\ud \sigma}=-y\psi\Big(\frac{6}{\kappa} - \ve6\xi(1-6\xi)\psi^{2}\Big) +
\ve(1-6\xi)\psi y^{2}(y+6\xi\psi)- \nonumber \\ &-\Big(\frac{6}{\kappa} -
\ve\big[(1-6\xi)y^{2} + 6\xi(y+\psi)^{2}\big]\Big)
\Big(\ve\frac{6}{\kappa}\big(1-\ve\kappa\xi\psi(y+\psi)\big) +
2\psi\big(y+6\xi\psi\big)\Big).
\end{align}
\end{subequations}
where we have made the following ``time'' transformation
\begin{equation}
\alpha\frac{\ud}{\ud\sigma}=\psi\Big(\frac{6}{\kappa} -
\ve6\xi(1-6\xi)\psi^{2}\Big) \frac{\ud}{\ud\ln{(a)}}
\label{eq:13}
\end{equation}
where the parameter $\alpha$ 
\begin{subequations}
\begin{align}
& \alpha =+1 \iff \psi\big(\frac{6}{\kappa} - \ve6\xi(1-6\xi)\psi^{2}\big)>0,\\
&\alpha=-1 \iff \psi\big(\frac{6}{\kappa} - \ve6\xi(1-6\xi)\psi^{2}\big)<0.
\end{align}
\end{subequations}
was introduced to preserve the orientation of the trajectories in this way that
on all of the phase portraits direction of arrows indicate the direction of
growth of the scale factor.

\begin{figure}
a)\includegraphics[scale=1]{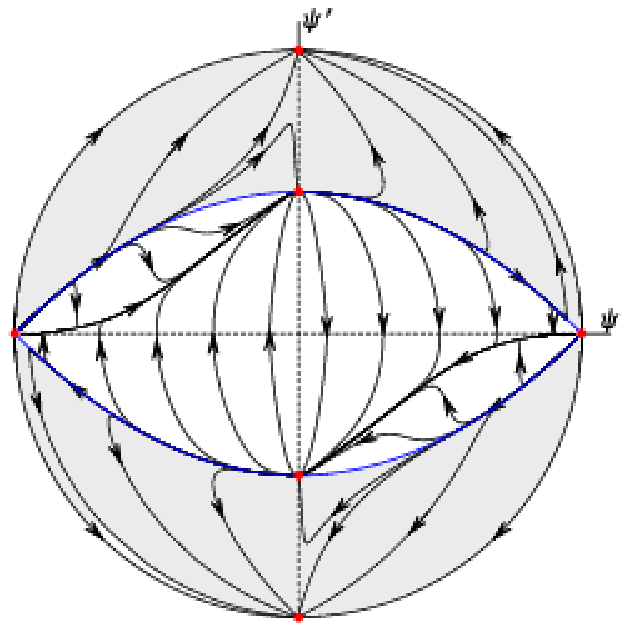} \\
b)\includegraphics[scale=1]{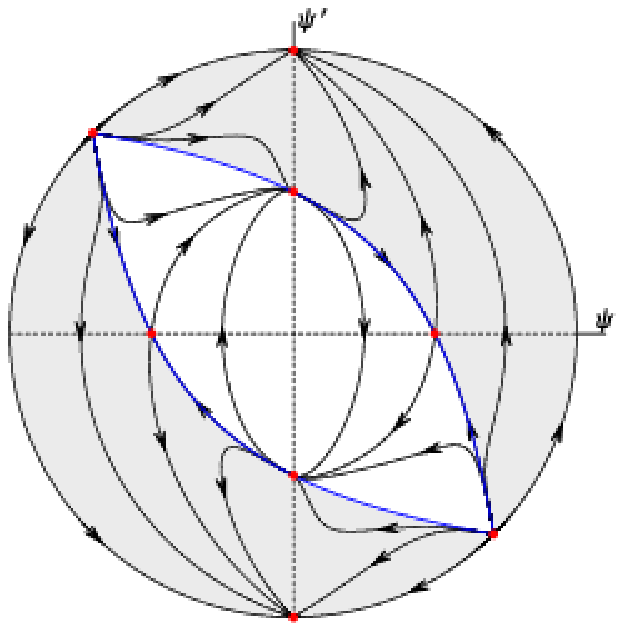}
\caption{The phase portraits for the canonical scalar field $\ve=1$ for: a)
minimal $\xi=0$ and b) conformal $\xi=1/6$ coupling. The shaded region denotes
nonphysical part of the phase space for the strictly positive potential
function. If the potential function is strictly negative the meaning of the
regions is reversed. The shape of the border between the regions does not depend
on the shape of the potential function. At the border of the physical region we
have two symmetric critical points at the $\psi$ axis for both cases. The value
of $H^{2}$ at that points is finite. The presence of a saddle type critical
point in the case b) is the effect of non-zero $\xi$.}
\label{fig:1}
\end{figure}

\begin{figure}
\includegraphics[scale=1]{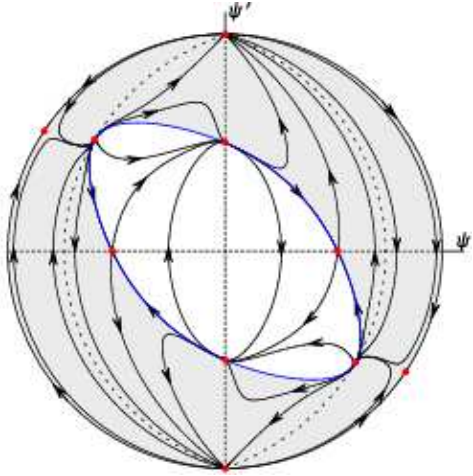}
\caption{The phase portrait for the canonical scalar field $\ve=1$ and
coupling constant $0<\xi<1/6$. The shaded region is nonphysical: 
$H^{2}<0$ for $V(\psi)>0$. There are three types of critical points at the
finite domain of the phase space: 1) $\psi_{0}=0$, $y_{0}^{2}=1$ and
$H^{2}=\text{const}.$ which are of
a stable or unstable node type; 2) $\psi_{0}^{2}=1/6\xi$, $y_{0}=0$,
$H^{2}=\infty$ of a saddle type 3) $\psi_{0}\ne 0$, $y_{0}\ne 0$,
$H^{2}=\infty$ of unstable node type for $V(\psi)>0$ and stable node type for
$V(\psi)<0$ (shaded region). The dashed line denotes singularity of ``time''
transformation (\ref{eq:13}). In comparison with the phase portrait
from Fig.~\ref{fig:1} for conformal coupling we can see that both phase
portraits are equivalent at the physical domain.}
\label{fig:2}
\end{figure}

\begin{figure}
a)\includegraphics[scale=1]{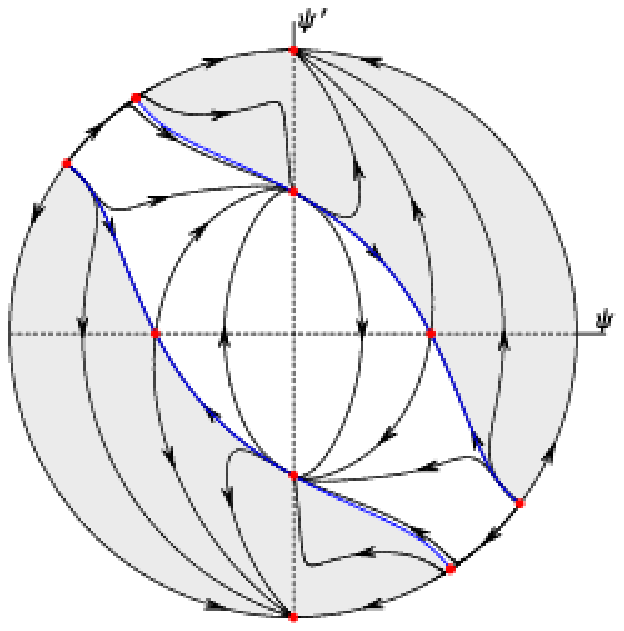} \\
b)\includegraphics[scale=1]{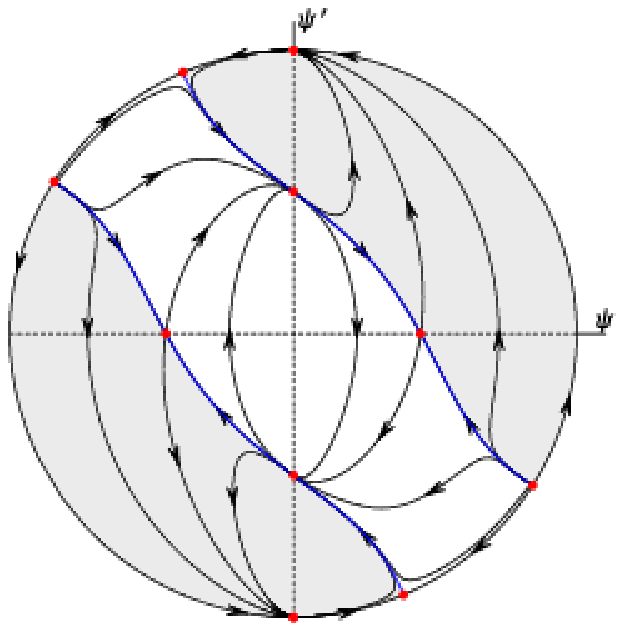} \\
c)\includegraphics[scale=1]{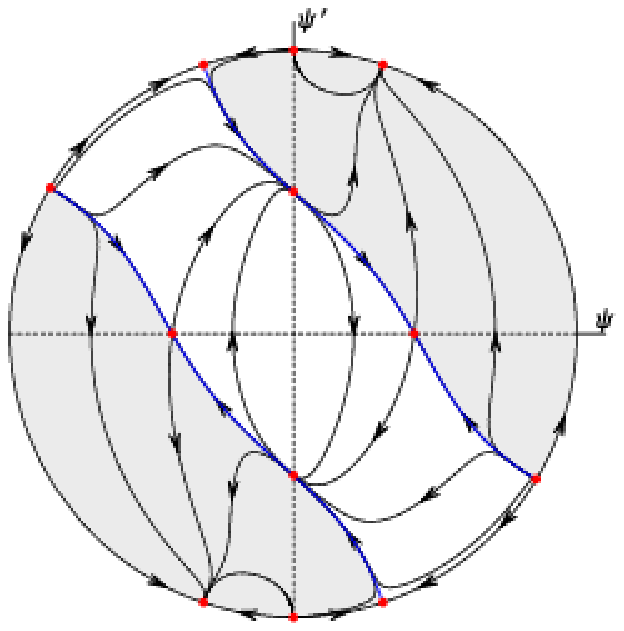}
\caption{The phase portraits for the canonical scalar
field $\ve=1$ and for the specific values of coupling constant: a) $\xi=3/16$,
b) $\xi=1/4$, c) $\xi=3/10$. In the cases a) and b) there are the critical
points at infinity of a mixed type (multiple critical points). At the
physical domain the phase portraits are topologically equivalent.}
\label{fig:3}
\end{figure}

\begin{figure}
a)\includegraphics[scale=1]{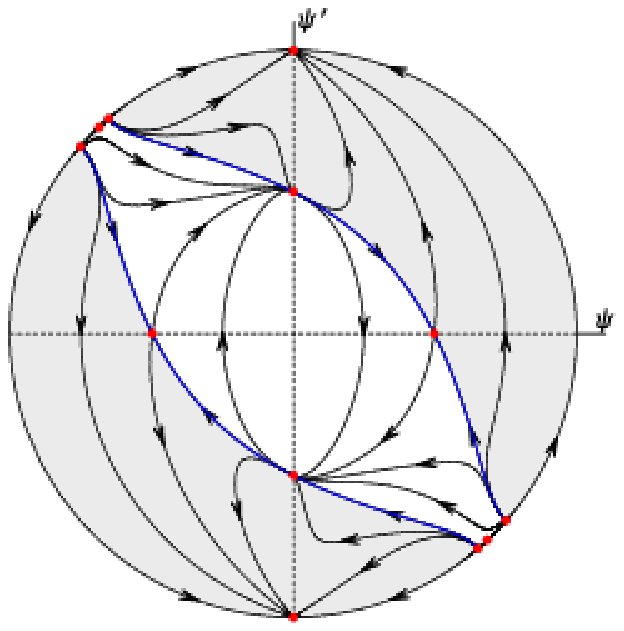} \\
b)\includegraphics[scale=1]{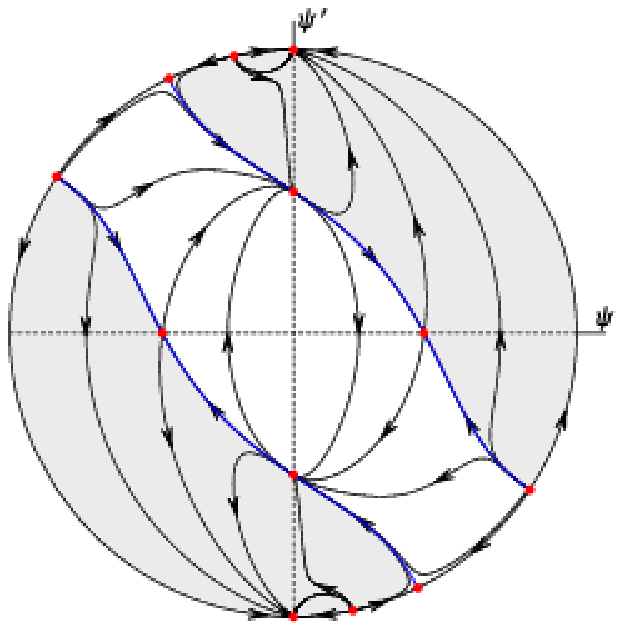} \\
c)\includegraphics[scale=1]{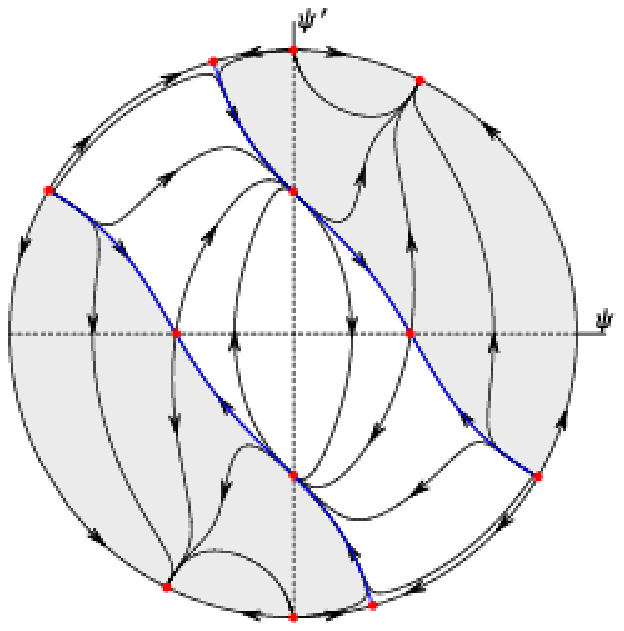}
\caption{The phase portraits for the canonical scalar field $\ve=1$ and
values of coupling constant: a) $1/6<\xi<3/16$, b) $3/16<\xi<1/4$, c)
$\xi=1/3$.}
\label{fig:4}
\end{figure}

\begin{figure}
a)\includegraphics[scale=1]{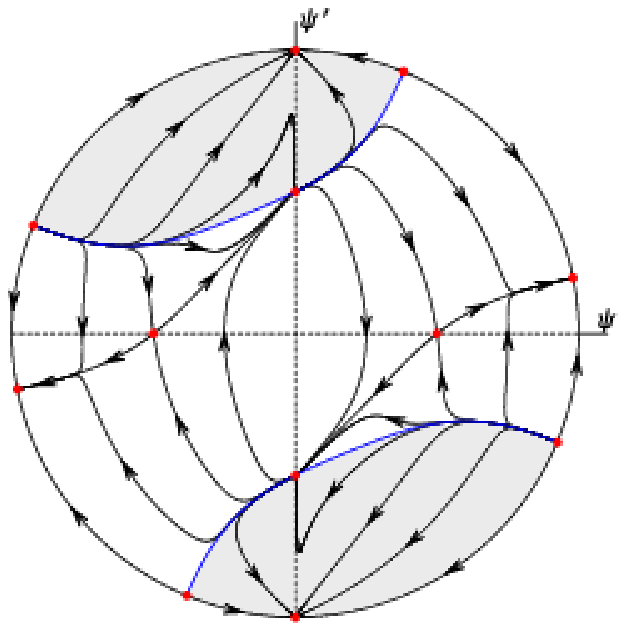} \\
b)\includegraphics[scale=1]{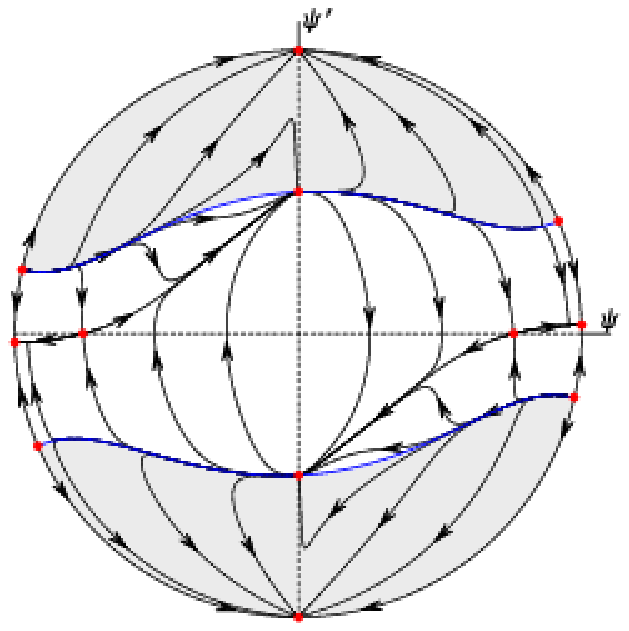}
\caption{The phase portraits for the canonical scalar field $\ve=1$ and
negative coupling constant $\xi<0$: in figure b) $\xi$ is greater, but
still negative, than in figure a). In the limit
$\xi\to0^{-}$ we receive phase portrait in Fig.~\ref{fig:1}a.}
\label{fig:5}
\end{figure}

\begin{figure}
a)\includegraphics[scale=1]{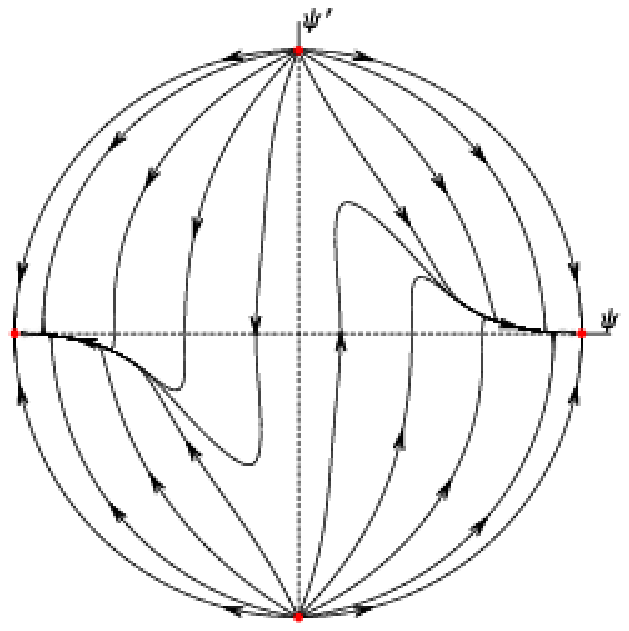} \\
b)\includegraphics[scale=1]{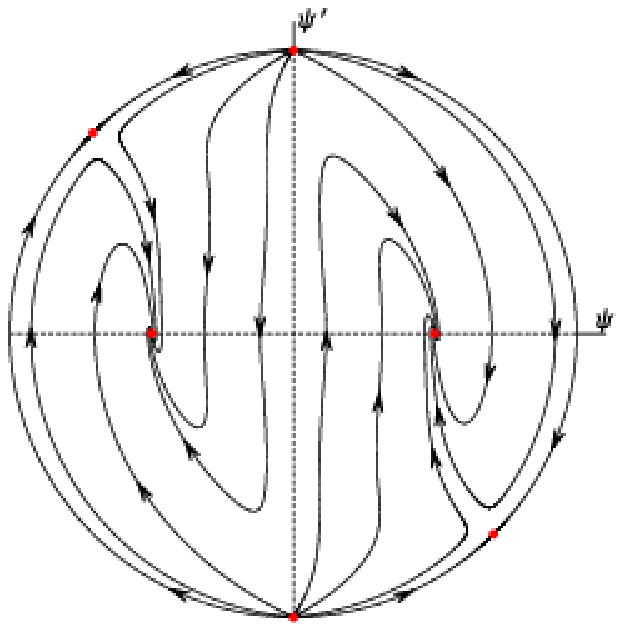}
\caption{The phase portraits for the phantom scalar field $\ve=-1$ and: a)
minimal
$\xi=0$ and b) conformal $\xi=1/6$ coupling. All the phase space $(\psi,\psi')$
is admissible only for the positive potential function. We can conclude, that
for
negative potential functions in the case of minimally or conformally coupled
phantom scalar fields, the scale factor is not a monotonic function of the
cosmological time. For the case b) a global attractor represents the de Sitter
state with $w_{\psi}=-1$. There are two types of trajectories which tend to
this attractor: 1) trajectories starting from $\psi=0$, $\psi'=\pm\infty$ state,
and 2) two single trajectories representing a separatrix of saddles at infinity (not shown).}
\label{fig:6}
\end{figure}

\begin{figure}
a)\includegraphics[scale=1]{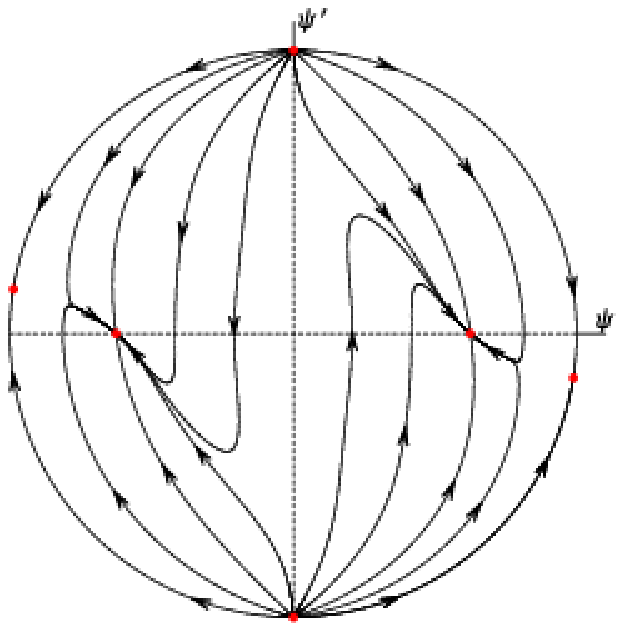} \\
b)\includegraphics[scale=1]{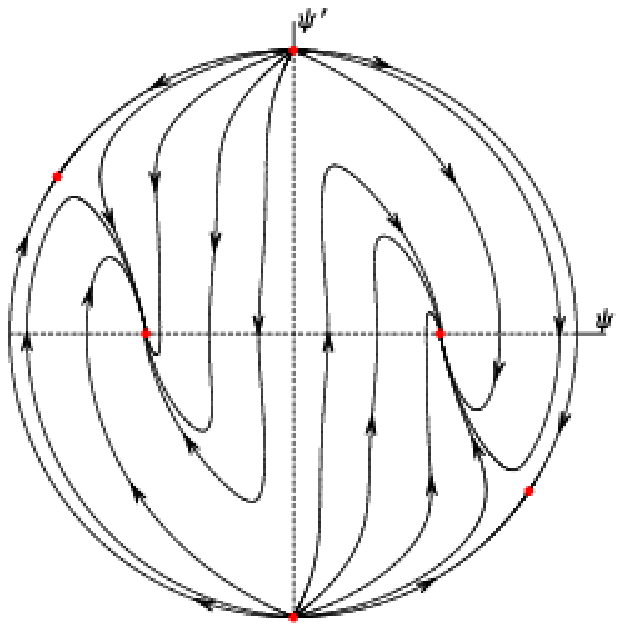}
\caption{The global phase portraits for the phantom scalar field and values of
coupling constant: a) $0<\xi<3/25$ and b) $3/25<\xi<1/6$. In the case a) in the
finite domain the critical domain is of a stable node type and in the case b) of
a focus type.}
\label{fig:7}
\end{figure}

\begin{figure}
a)\includegraphics[scale=1]{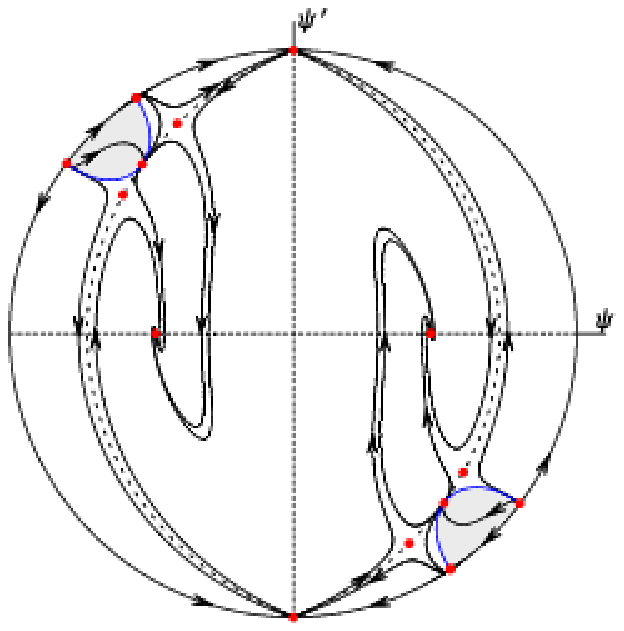} \\
b)\includegraphics[scale=1]{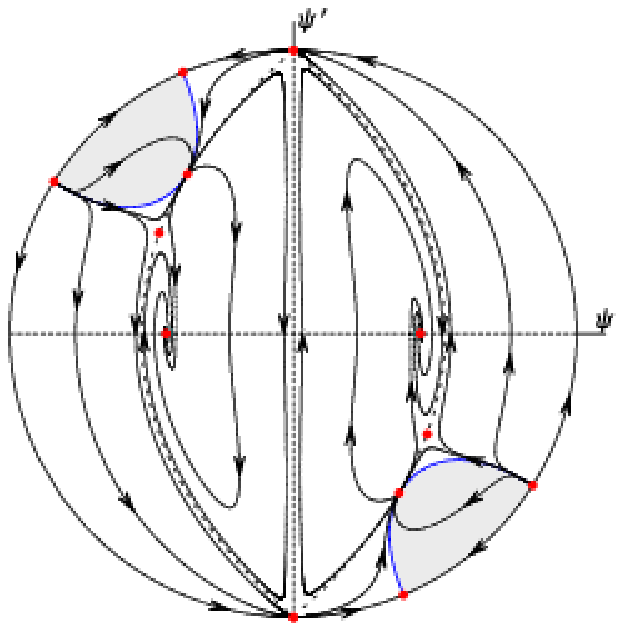} \\
c)\includegraphics[scale=1]{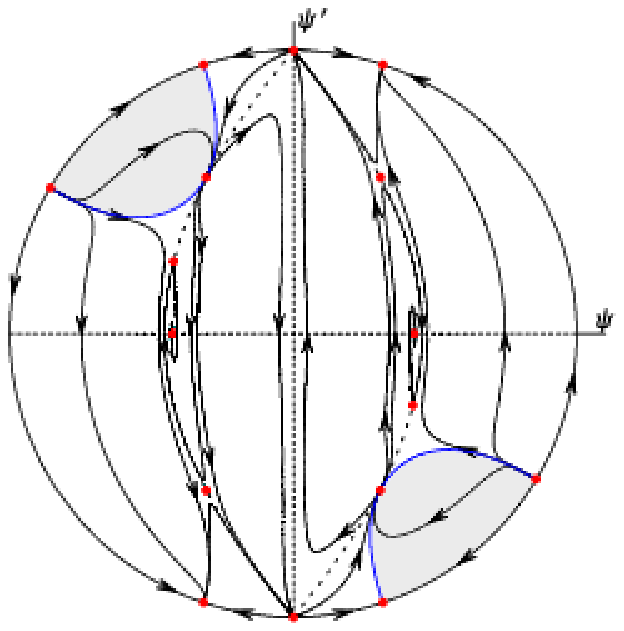}
\caption{The global phase portraits for the phantom scalar field
$\ve=-1$ and for the specific values of coupling constant: a) $\xi=3/16$, b)
$\xi=1/4$, c) $\xi=3/10$. In cases a) and b) one of the critical points at
infinity is of a mixed type (multiple critical points) (see Fig.~\ref{fig:3}). On
all figures one can see trajectories starting from the unstable node and landing
at the stable focus as a generic scenario of route to the de Sitter state.}
\label{fig:8}
\end{figure}

\begin{figure}
\includegraphics[scale=1]{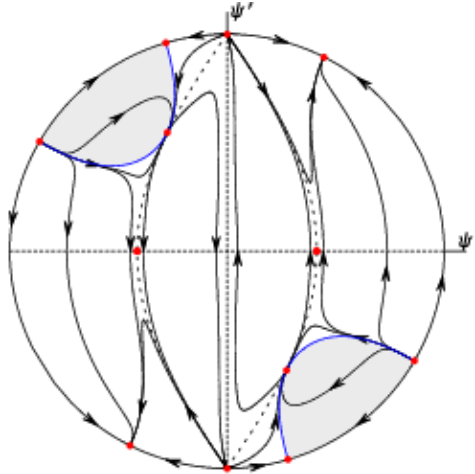}
\caption{The global phase portrait for the phantom scalar field $\ve=-1$ and
distinguished value of coupling constant $\xi=1/3$. In this case critical point
at the finite domain of the phase space is located at the line of singularities of
the time transformation (\ref{eq:13}).}
\label{fig:9}
\end{figure}

\begin{figure}
\includegraphics[scale=1]{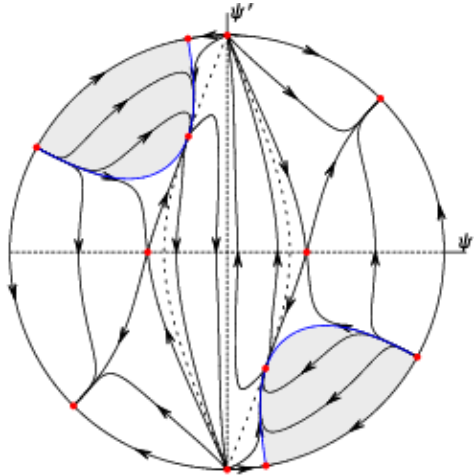}
\caption{The global phase portrait for the phantom scalar field $\ve=-1$ and the
values of coupling constant $\xi>1/3$. The characteristic critical point of a
focus type disappeared.}
\label{fig:10}
\end{figure}

\begin{figure}
a)\includegraphics[scale=1]{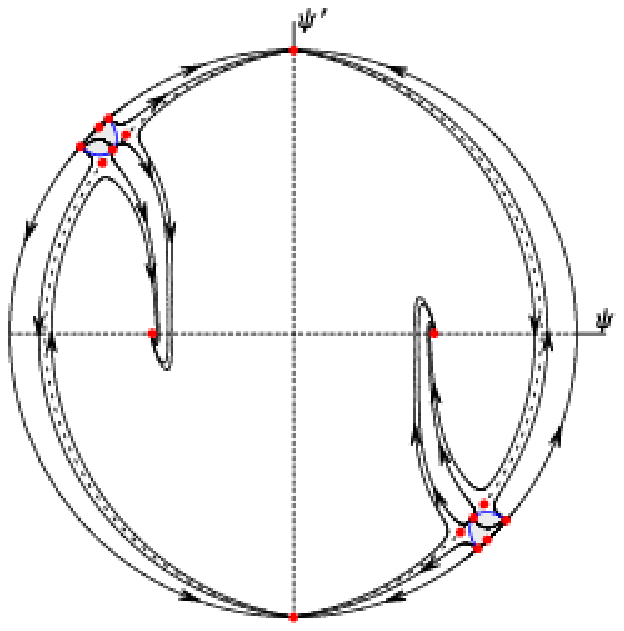} \\
b)\includegraphics[scale=1]{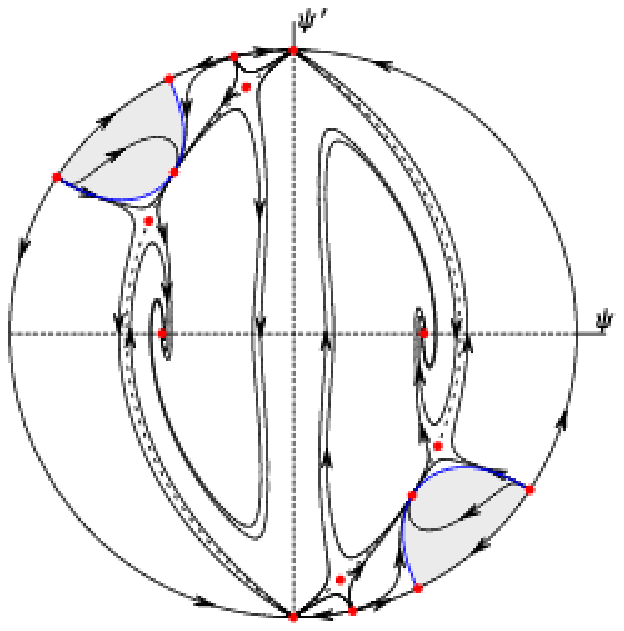} \\
c)\includegraphics[scale=1]{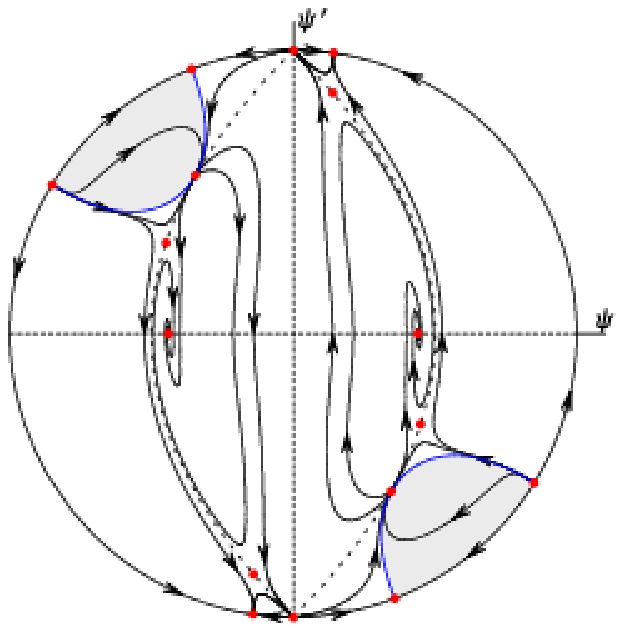}
\caption{The global phase portraits for the phantom scalar field $\ve=-1$ and
the values of coupling constant: a) $1/6<\xi<3/16$, b) $3/16<\xi<1/4$, c)
$1/4<\xi<3/10$. In all cases there is present scenario of reaching the global
attractor (a focus type critical point) from the unstable node. Note that in the
case c) not all of the trajectories starting from an unstable node are reaching
the de Sitter state, in contrast to cases a) and b).}
\label{fig:11}
\end{figure}

\begin{figure}
a)\includegraphics[scale=1]{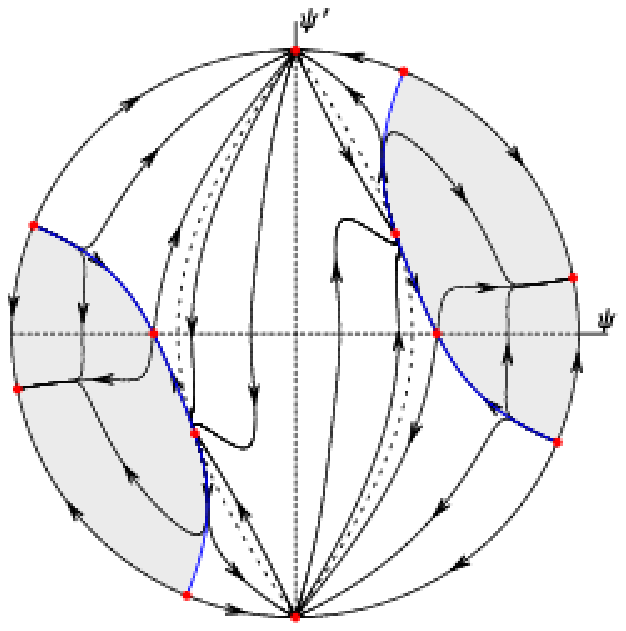} \\
b)\includegraphics[scale=1]{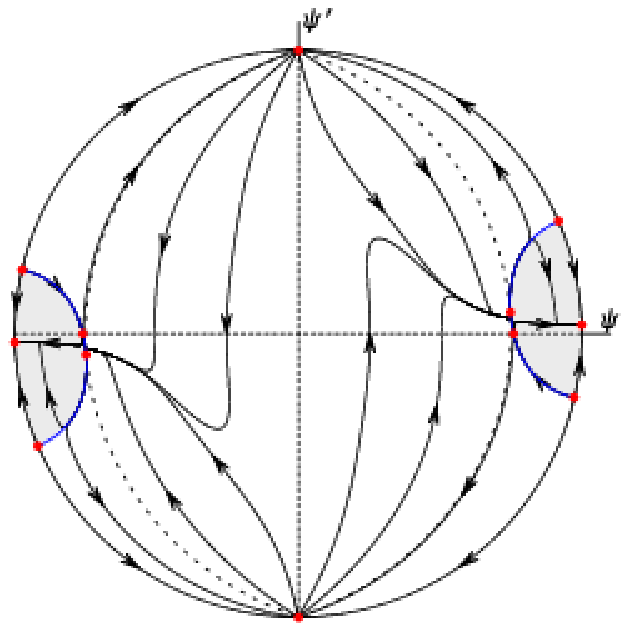}
\caption{The global phase portraits for the phantom scalar field $\ve=-1$ and
negative coupling constant $\xi<0$: in figure b) $\xi$ is greater, but still
negative, than in figure a). It is easy to notice that in the limit
$\xi\to0^{-}$ we receive phase portrait from the Fig.~\ref{fig:6}a. for the
phantom scalar field with minimal coupling.}
\label{fig:12}
\end{figure}

\begin{table*}
\caption{\label{tab:1} Finite critical points and their characters.}
\begin{ruledtabular}
\begin{tabular}{c|cc|c}
 &\multicolumn{2}{c|}{existence}\\
 & $\ve=+1$ & $\ve=-1$& eigenvalues \\ \hline
 $ \psi_{0}^{2}=\big(\frac{6}{\kappa}\big)\frac{1}{\ve6\xi}$, $y_{0}=0$ & $\xi>0$ & $\xi<0$ &
 $\lambda_{1}=\frac{1}{2}\big(\frac{6}{\kappa}\big)^{2}(\ve3-5)\frac{1}{\psi_{0}}$,
 $\lambda_{2}=\frac{1}{2}\big(\frac{6}{\kappa}\big)^{2}(\ve3+5)\frac{1}{\psi_{0}}$ \\
$\psi_{0}^{2}=-\big(\frac{6}{\kappa}\big)\frac{1}{\ve6\xi}$, $y_{0}=0$ & $\xi<0$ & $\xi>0$ &
$\lambda_{1,2}=-\big(\frac{6}{\kappa}\big)3(1-3\xi)\psi_{0}\pm\big(\frac{6}{\kappa}\big)^{3/2}\sqrt{-\frac{\ve}{2\xi}(1-3\xi)(3-25\xi)}$
\\
$\psi_{0}=0$, $y_{0}^{2}=\ve\frac{6}{\kappa}$ & $\forall \xi\in\mathbf{R}$ & -- &
$\lambda_{1}=\big(\frac{6}{\kappa}\big)y_{0}$,
$\lambda_{2}=\big(\frac{6}{\kappa}\big)2y_{0}$ \\
$\psi_{0}^{2}=\big(\frac{6}{\kappa}\big)\frac{1}{\ve6\xi(1-6\xi)}$,
$y_{0}=-\ve\frac{\frac{6}{\kappa}}{1-6\xi}\frac{1}{\psi_{0}}$ & $0<\xi<\frac{1}{6}$ & $\xi<0$
or $\xi>\frac{1}{6}$ & $\lambda_{1}=-\big(\frac{6}{\kappa}\big)2y_{0}$,
$\lambda_{2}=\big(\frac{6}{\kappa}\big)6\xi\psi_{0}$\\
$\psi_{0}^{2}=\big(\frac{6}{\kappa}\big)\frac{1}{\ve6\xi(1-6\xi)}$, & -- &
$\frac{1}{6}<\xi<\frac{1}{3}$& $\lambda_{1}=-\big(\frac{6}{\kappa}\big)2y_{0}$,
$\lambda_{2}=\big(\frac{6}{\kappa}\big)2y_{0}$\\
$y_{0}=-4\xi\frac{1-3\xi}{1-4\xi}\psi_{0}\pm
\frac{2\sqrt{-\ve3\xi(1-3\xi)\frac{6}{\kappa}}}{3(1-4\xi)}$
& & & \\

\end{tabular}
\end{ruledtabular}
\end{table*}

\section{Phase space analysis of dynamics}
\label{sec:3}

In this section we present detailed discussion of the type of critical
points of the dynamical system (\ref{syst}). 

If we write the evolutional equations for the non-minimally coupled scalar
field in the form of the dynamical system, the first step would be an
identification of the
critical points of the system. Physically they represent asymptotic (or
stationary) states of the system under considerations and mathematically
correspond to vanishing r.h.s. of the system. The second step is a
characterization of
the type of critical point which can be performed after calculation of the
eigenvalues of the linearization matrix calculated of this critical point. The
critical points are usually represented by physically interesting solutions and
these solutions can be attractors for trajectories in its neighborhood which
evolve to it independently on the initial conditions. In the quintessence
cosmology we
are looking for the attractors, which give rise to solutions with desired
property but we would like to know whether it is typical (generic) solution or
exceptional (non-generic). This is a reason of our interest in the stability
of the critical points.

For full dynamical analysis investigation of the behavior of the trajectories
at infinity is needed. It can be performed by construction of Poincar{\'e}
sphere \cite{Perko:1991}. If we complete the phase plane by a circle at infinity
which is a projection on the equator, then we receive the global phase portrait
with a circle at infinity. In our case r.h.s. contain the non-minimal coupling
constant $\xi$ as a free parameter. The global phase portraits depend on the value
of this parameter but for some ranges of the values of $\xi$ phase portraits can
be equivalent (indistinguishable from the dynamical system point of view). If we
fix the value of non-minimal coupling, then one can study the influence of this
parameter on the global dynamics. The equivalence of the phase portraits is
established by homeomorphism preserving direction of time along the
trajectories.

Critical point: $\psi_{0}=0$, $y_{0}^{2}=\ve$ exists only for the canonical
scalar field $\ve=+1$. Direct calculation of the Hubble function (\ref{hub}) at
this point gives an undefined symbol $\frac{0}{0}$. It is why we use the
linearized solutions in the vicinity of this critical point to show that the
Hubble function at this point is finite and depends on the initial conditions of
the linearized solutions. They are in the form
\begin{subequations}
\begin{align}
\psi(\sigma) & = \psi_{\text{(i)}}\exp{(\alpha\lambda_{1}\sigma)}, \\
y(\sigma) & = y_{0} -6\xi\psi_{\text{(i)}}\exp{(\alpha\lambda_{1}\sigma)} +
\big(6\xi\psi_{\text{(i)}}+(y_{\text{(i)}}-y_{0})\big)\exp{(\alpha\lambda_{2}\sigma)},
\end{align}
\end{subequations}
where $\lambda_{1}=y_{0}$ and $\lambda_{2}=2y_{0}$ are eigenvalues of the
linearization matrix calculated at this critical point, $\psi_{\text{(i)}}$ and
$y_{\text{(i)}}$ are initial conditions and $y_{0}$ is a coordinate of the
critical point.

Inserting those solution into the formula (\ref{hub}) we receive that the Hubble
function in the vicinity of the critical point ($\psi_{0}=0$, $y_{0}^{2}=\ve$)
is 
\begin{align}
H^{2}_{\text{lin}} = m^{2}\psi_{\text{(i)}}^{2}\exp{(2\alpha\lambda_{1}\sigma)}
\Big[& -6\xi(1-6\xi)\psi^{2}_{\text{(i)}}\exp{(2\alpha\lambda_{1}\sigma)} - 2
y_{0}\big(6\xi\psi_{\text{(i)}}+(y_{\text{(i)}}-y_{0})\big)\exp{(\alpha\lambda_{2}\sigma)}
- \nonumber \\
 &-
 \big(36\xi^{2}\psi_{(i)}^{2}+12\xi\psi_{(i)}(y_{(i)}-y_{0})+(y_{(i)}-y_{0})^{2}\big)\exp{(2\alpha\lambda_{2}\sigma)}\Big]^{-1}
\end{align}
then taking the limit value of this function for $\sigma\to\pm\infty$ (depending
on the critical point $y_{0}=\mp1$) we receive
\begin{equation}
\lim  H^{2}_{\text{lin}} = m^{2}\frac{\psi_{(i)}^{2}}
{-2y_{0}\big(6\xi\psi_{(i)}+(y_{(i)}-y_{0})\big)-6\xi(1-6\xi)\psi_{(i)}^{2}}
\end{equation}
which is always a positive quantity. For the special values of minimal $\xi=0$
(see Fig.~\ref{fig:1}a) and conformal coupling $\xi=1/6$ (see
Fig.~\ref{fig:1}b) the values of the Hubble function are
\begin{align}
H^{2}_{\text{lin}} &= m^{2}\frac{\psi_{(i)}^{2}}{-2y_{0}(y_{i}-y_{0})},
\qquad \qquad \qquad \text{for} \qquad \xi=0, \nonumber \\
H^{2}_{\text{lin}} &=
m^{2}\frac{\psi_{(i)}^{2}}{-2y_{0}\big(\psi_{(i)}+(y_{(i)}-y_{0})\big)},
\qquad \text{for} \qquad \xi=\frac{1}{6}. \nonumber
\end{align}

Critical point: $\psi_{0}^{2}=\frac{1}{\ve 6 \xi(1-6\xi)}$,
$y_{0}=-\ve\frac{1}{(1-6\xi)\psi_{0}}$ is very interesting because the Hubble
function (\ref{hub}) at this point is singular $H^{2}=\infty$ and
$\dot{H}=(\frac{1}{2}H^{2})'=\infty$. 

Linearized solutions in the vicinity of this critical points are
\begin{subequations}
\begin{align}
\psi(\sigma) & = \psi_{0} +
\exp{(\alpha\lambda_{2}\sigma)}\big(\psi_{(i)}-\psi_{0}\big),\\
y(\sigma) & = y_{0} + \exp{(\alpha\lambda_{1}\sigma)}
\Big(2(1-3\xi)\big(\psi_{(i)}-\psi_{0}\big) + \big(y_{(i)}-y_{0}\big)\Big) -
\exp{(\alpha\lambda_{2}\sigma)}\Big(2(1-3\xi)\big(\psi_{(i)}-\psi_{0}\big)\Big)
\end{align}
\end{subequations}
where $\lambda_{1}=-y_{0}$ and $\lambda_{2}=-2 y_{0}$ are
eigenvalues of the linearization matrix at the critical point, $\psi_{(i)}$ and
$y_{(i)}$ are initial conditions and $\psi_{0}$ and $y_{0}$ are coordinates of
the critical point.

Using time transformation (\ref{eq:13}) we can calculate the scale factor growth
along the trajectory
\begin{equation}
\Delta \ln{a} =
\int_{0}^{\infty}\psi(\sigma)\big(1-\ve6\xi(1-6\xi)\psi(\sigma)^{2}\big)
\ud\sigma,
\end{equation}
and the cosmological time growth
\begin{equation}
\Delta t = \int_{a_{i}}^{a_{f}} \frac{1}{H}\ud\ln{a} =
\int_{0}^{\infty}\frac{1}{H}\psi(\sigma)\big(1-\ve6\xi(1-6\xi)\psi(\sigma)^{2}\big)
\ud\sigma.
\end{equation}
Linearized solutions are good approximations of the original system in the
vicinity of the critical point. In what follows we assume that
$\big(\psi_{(i)}-\psi_{0}\big)^{2}=\big(y_{(i)}-y_{0}\big)^{2}=\big(\psi_{(i)}-\psi_{0}\big)\big(y_{(i)}-y_{0}\big)=0$,
$\alpha=1$ and $y_{0}>0$ (see Fig.~\ref{fig:12}). Then 
\begin{equation}
\Delta \ln{a} = \ve (1-6\xi)\big(\psi_{(i)}-\psi_{0}\big)\psi_{0}
\end{equation}
and
\begin{align}
\Delta t & = - \frac{1}{\sqrt{m^{2}}}\ve 12 \xi (1-6\xi)\big(\psi_{(i)}-\psi_{0}\big)
\psi_{0} \int_{0}^{\infty}\sqrt{A \exp{(-y_{0}\sigma)} + B \exp{(-2y_{0}\sigma)}}
\exp{(-2y_{0}\sigma)} \ud \sigma \nonumber \\ 
& = \frac{1}{\sqrt{m^{2}}} \ve 12 \xi(1-6\xi)\big(\psi_{(i)}-\psi_{0}\big)
\frac{1}{y_{0}24 B^{5/2}} \Big\{ \frac{3}{2}A^{2}\Big(\log{A} - 2
\log{(\sqrt{B}+\sqrt{A+B})}\Big) - \nonumber \\ & -
\sqrt{B(A+B)}(-A+2B)(3A+4B)\Big\}
\end{align}
where $A$ and $B$ are positive constants 
$$A=-\ve\big(y_{0}+12\xi\psi_{0}\big)\big(2(1-3\xi)(\psi_{(i)}-\psi_{0}) +
(y_{(i)}-y_{0})\big)$$
$$B=\ve4(1-6\xi)(y_{0}+3\xi\psi_{0})(\psi_{(i)}-\psi_{0}) $$
For every case of existence of such a critical point (see
Figs \ref{fig:2}, \ref{fig:8}, \ref{fig:9}, \ref{fig:10}, \ref{fig:11} for
an unstable node and Fig.~\ref{fig:12} for a stable node in the ``physical
region'') growth of the scale factor and the cosmological time is finite.

Now we calculate the first derivative of the Hubble function
(\ref{hub}) with respect to the cosmological time at this point
\begin{equation}
\dot{H} = \frac{1}{2}(H^{2})' = \Big(\frac{V(\psi)}{1-\ve[\psi'^{2} +
12\xi\psi\psi'+6\xi\psi^{2}]}\Big)'
\end{equation}
where a dot denotes differentiation with respect to the cosmological time and 
a prime with respect to the natural logarithm of the scale factor. Then after
the elimination of second derivative of the field with respect to the natural
logarithm of the scale factor we have
\begin{equation}
\dot{H} = -\frac{6\xi V'(\psi)\psi}{1-\ve6\xi(1-6\xi)\psi^{2}} - 
\frac{\ve2V(\psi)
\big((1-6\xi)\psi'^{2}+(\psi'+6\xi\psi)^{2}\big)}
{\big(1-\ve6\xi(1-6\xi)\psi^{2}\big) \big(1-\ve[\psi'^{2} +
12\xi\psi\psi'+6\xi\psi^{2}]\big)}
\end{equation}
This expression at the critical point $\psi_{0}^{2}=\frac{1}{\ve6\xi(1-6\xi)}$,
$\psi'_{0}=-\ve\frac{1}{1-6\xi}\frac{1}{\psi_{0}}$ is singular since the
numerator is finite and the denominator is equal zero.

The trajectories starting form this critical point corresponding to the
singularities of finite scale factor seems to be most interesting. For such
state appearing for both, canonical and phantom scalar fields we have curvature
singularity because Hubble parameter is infinite but the scale factor assumes
finite value. They are typical because the critical point is an unstable node
(see Fig.~\ref{fig:2} for an unstable node for the canonical scalar field and
and Figs.~\ref{fig:8}, \ref{fig:9}, \ref{fig:10} and \ref{fig:11} for an
unstable node and Fig.~\ref{fig:12} for a stable node for the phantom scalar
field).

Next we can proceed to the analysis of behaviour of system (\ref{syst}) at
the circle at infinity. Introducing the polar variables in order to compactify
the phase space by adjoining the circle at infinity
$$
\psi = \frac{r}{1-r}\cos{\theta}, \qquad y = \frac{r}{1-r}\sin{\theta},
$$
where $r$ and $\theta$ are the polar system coordinates, we receive the
following dynamical system
\begin{subequations}
\label{syscomp}
\begin{align}
\alpha \frac{\ud r}{\ud \sigma} r (1-r)^{3} & = r(1-r)\sin{\theta} 
\Bigg\{
r^{2}\cos{\theta}(\cos{\theta}-\sin{\theta})
\Big(\frac{6}{\kappa}(1-r)^{2} -\ve6\xi(1-6\xi)r^{2}\cos^{2}{\theta}\Big) +
\nonumber \\
& \qquad \qquad \qquad  + \ve(1-6\xi)r^{4}\sin^{2}{\theta}\cos{\theta}(\sin{\theta}+6\xi\cos{\theta})
- \nonumber \\
& \qquad \qquad \qquad - \Big(\frac{6}{\kappa}(1-r)^{2}-\ve r^{2}\big((1-6\xi)\sin^{2}{\theta} +
6\xi(\sin{\theta}+\cos{\theta})^{2}\big)\Big)\nonumber \\
& \qquad \Big(\ve\frac{6}{\kappa}\big((1-r)^{2}-\ve\kappa\xi
r^{2}\cos{\theta}(\sin{\theta}+\cos{\theta})\big) + 2 r^{2}
\cos{\theta}(\sin{\theta}+6\xi\cos{\theta})\Big)
\Bigg\}, \\
\alpha \frac{\ud \theta}{\ud \sigma} r(1-r)^{3} & = \cos{\theta}
\Bigg\{
-r^{2}\sin{\theta}(\sin{\theta}+\cos{\theta})
\Big(\frac{6}{\kappa}(1-r)^{2} -\ve6\xi(1-6\xi)r^{2}\cos^{2}{\theta}\Big) + 
\nonumber \\
& \qquad \qquad \qquad  +
\ve(1-6\xi)r^{4}\sin^{2}{\theta}\cos{\theta}(\sin{\theta}+6\xi\cos{\theta})
- \nonumber \\
& \qquad \qquad \qquad - \Big(\frac{6}{\kappa}(1-r)^{2}-\ve
r^{2}\big((1-6\xi)\sin^{2}{\theta} +
6\xi(\sin{\theta}+\cos{\theta})^{2}\big)\Big)\nonumber \\
& \qquad \Big(\ve\frac{6}{\kappa}\big((1-r)^{2}-\ve\kappa\xi
r^{2}\cos{\theta}(\sin{\theta}+\cos{\theta})\big) + 2 r^{2}
\cos{\theta}(\sin{\theta}+6\xi\cos{\theta})\Big)
\Bigg\}.
\end{align}
\end{subequations}

For identification of the critical points at infinity we can calculate the 
polar angle $\theta$ from those equations simply by putting $r=1$. Then we 
receive the following equation for the direction $\theta$ determining the 
localization of the critical points
\begin{equation}
\ve 3 \cos^{2}{\theta}\big(2\xi\cos{\theta}+(1-4\xi)\sin{\theta}\big)
\big(\sin^{2}{\theta}+6\xi\cos^{2}{\theta}+12\xi\sin{\theta}\cos{\theta}\big) =
0.
\end{equation}
We can notice that position of the critical points at infinity do not depend on
the form of scalar field assumed, i.e. critical points are the same for both
canonical and phantom scalar fields. In Table~\ref{tab:2} we have gathered
positions and the character of these points. Note that there are specific values
of the coupling constant $\xi$ for which some critical points coincide. For example
for minimal $\xi=0$ and conformal $\xi=\frac{1}{6}$ coupling critical points
$2)$ and $3)$ in Table \ref{tab:2} have the same location. This is
the reason why in these cases the critical points at infinity are degenerated
(i.e. the eigenvalues of linearization matrix calculated for the critical points 
located at infinity is identity zero). Such points are called multiple critical 
points. In our approach we treat the coupling constant $\xi$ as a control  
parameter for which the bifurcation analysis is performed. The values of $\xi$ different from
minimal and conformal coupling split multiple critical points and remove the
degeneration.

\begin{table*}
\caption{\label{tab:2} The critical points at infinity and their characters.}
\begin{ruledtabular}
\begin{tabular}{c|c|c|c}
 & Critical point & existence & eigenvalues \\
\hline
1) & $\cos{\theta}=0$ & $\forall \xi\in\mathbf{R}$ & $\lambda_{1}=\lambda_{2}=0$\\
2) & $\tan{\theta}= - \frac{2\xi}{1-4\xi}$ & $\forall
\xi\in\mathbf{R}\setminus\{\frac{1}{4}\}$ &
$\lambda_{1}=\ve6\xi(1-6\xi)\frac{3-16\xi}{1-4\xi}$,
$\lambda_{2}=-\ve12\xi^{2}\frac{1-6\xi}{1-4\xi}$\\
3) & $\tan{\theta}=-6\xi\pm\sqrt{6\xi(6\xi-1)}$ & $\xi \le 0$ or $\xi \ge
\frac{1}{6}$&
$\lambda_{1}=-\ve12\xi(1-6\xi)\big(3(1-4\xi)\mp2\sqrt{6\xi(6\xi-1)}\big)$,\\
& & & $\lambda_{2}=-\ve6\xi(1-6\xi)\Big(6\xi\pm\sqrt{6\xi(6\xi-1)}\Big)$ \\
\end{tabular}
\end{ruledtabular}
\end{table*}

Now we can simply calculate the value of the Hubble function (\ref{hub}) at
critical points at infinity. Using the polar coordinates with compactification
with circle at infinity we receive
\begin{equation}
H^{2}|_{\infty} = -\ve
m^{2}\frac{\cos^{2}{\theta}}{\sin^{2}{\theta}+12\xi\sin{\theta}\cos{\theta}+6\xi\cos^{2}{\theta}}.
\end{equation}
Inserting the position angle of the critical points at infinity to the above
expression we can conclude that at critical point $1)$ in Table \ref{tab:2}
$H^{2}=0$, for the second point $H^{2}=-\ve
m^{2}\frac{(1-4\xi)^{2}}{2\xi(1-6\xi)(3-16\xi)}$, and finally for the point
$3)$ the Hubble function is singular $H^{2}=\infty$. The most interesting critical point seems
to be the second critical point because for the wide range of values of
the parameter $\xi$ the final state can be the de Sitter attractor with $H^{2}=\text{const}$ 
in spite of that the phase variables $\psi$ and $\psi'$ assume the infinite values.

\section{Conclusions}

We study the dynamics of a scalar field with a simple quadratic potential
function and non-minimally coupled to the gravity via $\xi R \psi^{2}$ term,
where $R$ is the Ricci scalar of the Robertson-Walker spacetime. We reduce
dynamical problem to the autonomous dynamical system on the phase plane in the
variables $\psi$ and its derivative with respect to the natural logarithm of the
scale factor. The constraint condition is solved in such a way that the final
dynamical system is free from the constraint and is defined on the plane. We
investigate the whole dynamics at the finite domain and at infinity. All the
trajectories for all admissible initial conditions are classified, and critical
points representing the asymptotic states (stationary solutions) are found. We
explore generic evolutionary paths to find the stable de Sitter state as a
global attractor and classify typical routes to this point. We study the effects
of the canonical scalar field as well as the phantom scalar field. The following
conclusions, as the results of our studies, can be drawn:
\begin{itemize}
\item[1.]{The cosmological models with the quadratic potential function and 
non-minimal coupling term $\xi R \psi^{2}$ are represented by a 2-dimensional
autonomous dynamical system which is studied in details on the phase plane
$(\psi,\psi')$ (see Table \ref{tab:1} for the critical points at the finite
domain). The shape of the physical region $H^{2}\ge0$ does not depend on
the form of the potential function, but only on the value of the coupling constant
$\xi$;}
\item[2.]{We investigated the fixed points of the dynamical system and their
stability to find the generic evolutional scenarios. We have shown the
existence of a finite scale factor singular point (both future and past) where
the Hubble function as well as its first cosmological time derivative diverge
$H^{2}=\infty$ and $\dot{H}=\infty$;}
\item[3.]{For the phantom scalar field $\ve=-1$ we found existence of a sink type
critical point (i.e. a stable node or a focus, depending on the value of the 
parameter $\xi$) which represents the de Sitter solution. Only for $\xi<1/6$ the
evolutional paths avoid a past finite scale factor singularity (see for example
Figs~\ref{fig:6} and \ref{fig:8});}
\item[4.]{For the canonical scalar field $\ve=1$ we found that for $0<\xi<1/6$
exists the critical point which corresponds to a past finite scale factor
singularity for models with $V(\psi)=\frac{1}{2}m^{2}\psi^{2}>0$. If $m^{2}<0$
then this point corresponds to a future finite scale factor singularity (see
Fig.~\ref{fig:2}).}
\end{itemize}

\begin{acknowledgments}
OH is very grateful to the organizers of ``School and Workshop on Dynamical
Systems'' 20 June - 18 July 2008 and ``Summer School in Cosmology'' 21 July - 1
August 2008 held at the Abdus Salam ICTP (Miramare--Trieste, Italy) for
hospitality during those activities, where part of this publication was
prepared.
MS is very grateful to prof. Mauro Carfora for discussion and warm atmosphere 
during the visit in Pavia where this work has been prepared. 
This work has been supported by the Marie Curie Host Fellowships for the
Transfer of Knowledge project COCOS (Contract No. MTKD-CT-2004-517186).
\end{acknowledgments}

\end{document}